# Techniques for Enhanced Physical-Layer Security

Pedro C. Pinto, *Student Member, IEEE*, João Barros, *Member, IEEE*, and Moe Z. Win, *Fellow, IEEE*

*Abstract*—Information-theoretic security—widely accepted as the strictest notion of security—relies on channel coding techniques that exploit the inherent randomness of propagation channels to strengthen the security of communications systems. Within this paradigm, we explore strategies to improve secure connectivity in a wireless network. We first consider the *intrinsically secure communications graph* ($i\mathcal{S}$-graph), a convenient representation of the links that can be established with information-theoretic security on a large-scale network. We then propose and characterize two techniques—*sectorized transmission* and *eavesdropper neutralization*—which are shown to dramatically enhance the connectivity of the $i\mathcal{S}$-graph.

## I. INTRODUCTION

Wireless communication is particularly susceptible to eavesdropping due to the broadcast nature of the transmission medium. In current systems, security is addressed above the physical layer using cryptographic protocols (e.g., RSA and AES), which assume that an error-free physical link has already been established. In contrast with this paradigm, it is possible to take advantage of physical-layer techniques to significantly strengthen the security of communication systems. This is the basic principle of *information-theoretic security*, widely accepted as the strictest notion of security.[1] It relies on channel coding techniques that exploit the inherent randomness of propagation channels to ensure that the transmitted messages cannot be decoded by a malicious eavesdropper.

The basis for information-theoretic security, which builds on the notion of perfect secrecy [1], was laid in [2] and later in [3], [4]. More recently, there has been a renewed interest in information-theoretic security over wireless channels, from the perspective of space-time communications [5], multiple-input multiple-output communications [6]–[10], eavesdropper collusion [11], [12], cooperative relay networks [13], fading channels [14]–[18], strong secrecy [19], [20], secret key agreement [21]–[25], code design [26]–[28], among other topics. A fundamental limitation of this literature is that it only considers scenarios with a small number of nodes. To account for large-scale networks composed of multiple legitimate and eavesdropper nodes, *secrecy graphs* were introduced in [29] from a geometrical perspective, and in [30] from an information-theoretic perspective. The local connectivity of secrecy graphs was extensively characterized in [31], while the scaling laws of the secrecy capacity were presented in [32], [33]. The feasibility of long-range secure communication was proved in [34], in the context of continuum percolation.

In this paper, we explore strategies to improve secure connectivity in a wireless network. We first consider the *intrinsically secure communications graph* ($i\mathcal{S}$-graph)—a convenient representation of the links that can be established with information-theoretic security on a large network. We then propose two techniques for improving the connectivity of the $i\mathcal{S}$-graph: (i) *sectorized transmission*, whereby each legitimate node transmits independently in multiple sectors of the plane (e.g., using directional antennas); and (ii) *eavesdropper neutralization*, whereby each legitimate node guarantees the absence of eavesdroppers in a surrounding region (e.g., by deactivating such eavesdroppers). We quantify and compare the effectiveness of the proposed techniques in terms of the resulting average node degree.

This paper is organized as follows. Section II describes the system model. Section III proposes and compares techniques for enhancing the secrecy of communication. Section IV provides some numerical results. Section V summarizes important findings.

## II. SYSTEM MODEL

### A. Wireless Propagation

Given a transmitter node $x_i \in \mathbb{R}^d$ and a receiver node $x_j \in \mathbb{R}^d$, we model the received power $P_{\text{rx}}(x_i, x_j)$ associated with the wireless link $\overrightarrow{x_i x_j}$ as

$$P_{\text{rx}}(x_i, x_j) = P \cdot g(x_i, x_j),$$

where $P$ is the transmit power, and $g(x_i, x_j)$ is the power gain of the link $\overrightarrow{x_i x_j}$. The gain $g(x_i, x_j)$ is considered constant (quasi-static) throughout the use of the communications channel, corresponding to channels with a large coherence time. Furthermore, the function $g$ is assumed to satisfy the following conditions, which are typically observed in practice: i) $g(x_i, x_j)$ depends on $x_i$ and $x_j$ only through the link length $|x_i - x_j|$;[2] ii) $g(r)$ is continuous and strictly decreasing with $r$; and iii) $\lim_{r \to \infty} g(r) = 0$.

### B. $i\mathcal{S}$-Graph

Consider a wireless network where the legitimate nodes and the potential eavesdroppers are randomly scattered in

---

P. C. Pinto and M. Z. Win are with the Laboratory for Information and Decision Systems (LIDS), Massachusetts Institute of Technology, Room 32-D674, 77 Massachusetts Avenue, Cambridge, MA 02139, USA (e-mail: ppinto@alum.mit.edu, moewin@mit.edu). J. Barros is with Departamento de Engenharia Electrotécnica e de Computadores, Faculdade de Engenharia da Universidade do Porto, Portugal (e-mail: jbarros@fe.up.pt).

This research was supported, in part, by the MIT Institute for Soldier Nanotechnologies, the Office of Naval Research Presidential Early Career Award for Scientists and Engineers (PECASE) N00014-09-1-0435, and the National Science Foundation under grant ECCS-0901034.

[1]Information-theoretic security is also known as *physical-layer security*, or *intrinsic security*. In the literature, the term "security" typically encompasses three different characteristics: *secrecy* (or privacy), *integrity*, and *authenticity*. We do not consider the issues of integrity or authenticity, and use the terms "secrecy" and "security" interchangeably.

[2]With abuse of notation, we write $g(r) \triangleq g(x_i, x_j)|_{|x_i - x_j| \to r}$.

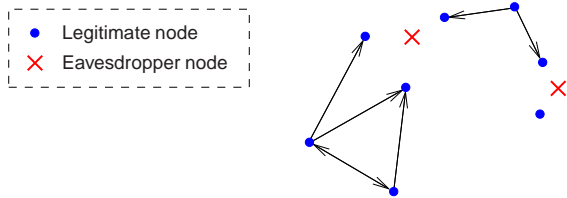

Figure 1. Example of an $i\mathcal{S}$-graph on $\mathbb{R}^2$.

space, according to some point processes. The $i\mathcal{S}$-graph is a convenient representation of the links that can be established with information-theoretic security on such a network.

*Definition 2.1 ($i\mathcal{S}$-Graph [31]):* Let $\Pi_\ell = \{x_i\}_{i=1}^\infty \subset \mathbb{R}^d$ denote the set of legitimate nodes, and $\Pi_e = \{e_i\}_{i=1}^\infty \subset \mathbb{R}^d$ denote the set of eavesdroppers. The $i\mathcal{S}$-graph is the directed graph $G = \{\Pi_\ell, \mathcal{E}\}$ with vertex set $\Pi_\ell$ and edge set

$$\mathcal{E} = \{\overrightarrow{x_i x_j} : \mathcal{R}_{\mathrm{s}}(x_i, x_j) > \varrho\}, \quad (1)$$

where $\varrho$ is a threshold representing the prescribed infimum secrecy rate for each communication link; and $\mathcal{R}_{\mathrm{s}}(x_i, x_j)$ is the *maximum secrecy rate* (MSR) of the link $\overrightarrow{x_i x_j}$, given by

$$\mathcal{R}_{\mathrm{s}}(x_i, x_j) = \left[ \log_2 \left(1 + \frac{P_{\mathrm{rx}}(x_i, x_j)}{\sigma_\ell^2}\right) - \log_2 \left(1 + \frac{P_{\mathrm{rx}}(x_i, e^*)}{\sigma_{\mathrm{e}}^2}\right) \right]^+$$

in bits per complex dimension, where $[x]^+ = \max\{x, 0\}$; $\sigma_\ell^2, \sigma_{\mathrm{e}}^2$ are the noise powers of the legitimate users and eavesdroppers, respectively; and $e^* = \underset{e_k \in \Pi_{\mathrm{e}}}{\mathrm{argmax}}\, P_{\mathrm{rx}}(x_i, e_k)$.[3]

The above definition admits an outage interpretation, in the sense that legitimate nodes set a target secrecy rate $\varrho$ at which they transmit without knowing the channel state information (CSI) of the legitimate nodes and eavesdroppers. In this context, an edge between two nodes in the $i\mathcal{S}$-graph signifies that the corresponding channel is not in secrecy outage.

In the remainder of the paper, we consider the particular scenario where the following conditions hold: (a) the noise powers of the legitimate users and eavesdroppers are equal, i.e., $\sigma_\ell^2 = \sigma_{\mathrm{e}}^2 = \sigma^2$; and (b) the infimum desired secrecy rate is zero, i.e., $\varrho = 0$.[4] Under these special conditions, the edge set in (1) simplifies to

$$\mathcal{E} = \left\{ \overrightarrow{x_i x_j} : |x_i - x_j| < |x_i - e^*|, \quad e^* = \underset{e_k \in \Pi_{\mathrm{e}}}{\mathrm{argmin}} |x_i - e_k| \right\},$$

which corresponds to the geometrical model proposed in [29]. Fig. 1 shows an example of such an $i\mathcal{S}$-graph on $\mathbb{R}^2$.

The spatial location of the legitimate and eavesdropper nodes can be modeled either deterministically or stochastically. In many cases, the node positions are unknown to the network designer a priori, so they may be treated as uniformly random according to a Poisson point process [35], [36].

---

[3]This definition uses *strong secrecy* as the condition for information-theoretic security. See [19], [31, Sec. II-B] for more details.

[4]Note that by setting $\varrho = 0$ we are considering the *existence* of secure links, in the sense that an edge $\overrightarrow{x_i x_j}$ is present iff $\mathcal{R}_{\mathrm{s}}(x_i, x_j) > 0$. The general case of non-zero secrecy rate threshold, $\varrho > 0$, and unequal noise powers, $\sigma_\ell^2 \neq \sigma_{\mathrm{e}}^2$, can still be considered (see [31, Sec. III-E] for an analysis) but does not provide as many insights.

*Definition 2.2 (Poisson $i\mathcal{S}$-Graph):* The Poisson $i\mathcal{S}$-graph is an $i\mathcal{S}$-graph where $\Pi_\ell, \Pi_{\mathrm{e}} \subset \mathbb{R}^d$ are mutually independent, homogeneous Poisson point processes with densities $\lambda_\ell$ and $\lambda_{\mathrm{e}}$, respectively.

In the remainder of the paper (unless otherwise indicated), we focus on Poisson $i\mathcal{S}$-graphs on $\mathbb{R}^2$.

## III. TECHNIQUES FOR ENHANCED SECURE COMMUNICATION

In this section, we first characterize the connectivity of the $i\mathcal{S}$-graph without any enhancement (as introduced in Definition 2.1), and then propose two techniques for improving secure connectivity: sectorized transmission and eavesdropper neutralization. For each strategy, we characterize the *average node degree* of a typical node in the corresponding enhanced $i\mathcal{S}$-graph.[5] The average degree is a measure of secure connectivity, and is used in this paper to quantify and compare the effectiveness of each strategy.

### A. Secure Communication Without Enhancement

We start by characterizing secure connectivity in the $i\mathcal{S}$-graph when no particular strategy is used.

*Theorem 3.1 (No Enhancement):* The out-degree $N_{\mathrm{out}}$ of a typical node in the Poisson $i\mathcal{S}$-graph has the following geometric probability mass function (PMF)

$$p_{N_{\mathrm{out}}}(n) = \left(\frac{\lambda_\ell}{\lambda_\ell + \lambda_{\mathrm{e}}}\right)^n \left(\frac{\lambda_{\mathrm{e}}}{\lambda_\ell + \lambda_{\mathrm{e}}}\right), \quad n \geq 0. \quad (2)$$

In particular, the average node degrees are given by

$$\mathbb{E}\{N_{\mathrm{out}}\} = \mathbb{E}\{N_{\mathrm{in}}\} = \frac{\lambda_\ell}{\lambda_{\mathrm{e}}}. \quad (3)$$

*Proof:* See [29], [30]. □

We observe that in the $i\mathcal{S}$-graph without enhancement, even a small density of eavesdroppers is enough to significantly disrupt connectivity. For example, if the density of eavesdroppers is half the density of legitimate nodes, then the average node degree is only $\frac{\lambda_\ell}{\lambda_{\mathrm{e}}} = 2$. In what follows, we propose techniques that achieve an average degree higher than $\frac{\lambda_\ell}{\lambda_{\mathrm{e}}}$.

### B. Secure Communication With Sectorized Transmission

We have so far assumed that the legitimate nodes employ omnidirectional antennas, distributing power equally among all directions in space. We now let each legitimate node transmit independently in $L$ sectors of the plane, with $L \geq 1$. This can be accomplished, for example, through the use of $L$ directional antennas. Our goal is to characterize the impact of the number of sectors $L$ on the local connectivity of the resulting $i\mathcal{S}$-graph.

With each node $x_i \in \Pi_\ell$, we associate $L$ transmission sectors $\{\mathcal{S}_i^{(l)}\}_{l=1}^L$, defined as

$$\mathcal{S}_i^{(l)} \triangleq \left\{ z \in \mathbb{R}^2 : \phi_i + (l-1)\frac{2\pi}{L} < \angle\overrightarrow{x_i z} < \phi_i + l\frac{2\pi}{L} \right\}$$

---

[5]We analyze the properties of a *typical node*, a notion which is made precise in [37, Sec. 4.4] using Palm theory. Specifically, Slivnyak's theorem states that the properties observed by a typical legitimate node $x \in \Pi_\ell$ are the same as those observed by node 0 in the process $\Pi_\ell \cup \{0\}$. Informally, a typical node of $\Pi_\ell$ is one that is uniformly picked from a finite region expanding to $\mathbb{R}^2$.

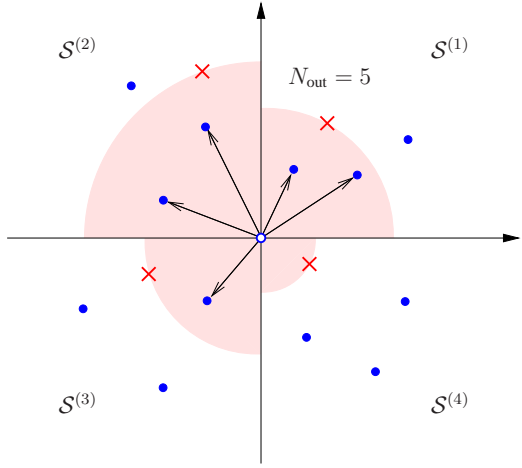

Figure 2. Secure communication with sectorized transmission. In this example with $L = 4$ sectors, the node at the origin can transmit messages with information-theoretic security to $N_{\text{out}} = 5$ nodes.

for $l = 1 \ldots L$, where $\{\phi_i\}_{i=1}^{\infty}$ are random offset angles with an arbitrary joint distribution. The resulting $i\mathcal{S}$-graph with sectorization, $G_L = \{\Pi_\ell, \mathcal{E}_L\}$, has an edge set given by

$$\mathcal{E}_L = \{\overrightarrow{x_i x_j} : |x_i - x_j| < |x_i - e^*|\},$$

where

$$e^* = \operatorname*{argmin}_{e_k \in \Pi_e \cap \mathcal{S}^*} |x_i - e_k|, \qquad \mathcal{S}^* = \left\{\mathcal{S}_i^{(l)} : x_j \in \mathcal{S}_i^{(l)}\right\}.$$

Here, $\mathcal{S}^*$ is the transmission sector of $x_i$ that contains the destination node $x_j$, and $e^*$ is the eavesdropper inside $\mathcal{S}^*$ that is closest to the transmitter $x_i$. Then, the secure link $\overrightarrow{x_i x_j}$ exists if and only if $x_j$ is closer to $x_i$ than any other eavesdropper inside the same transmission sector where the destination $x_j$ is located. The following theorem characterizes the distribution of the out-degree.

*Theorem 3.2 (Sectorized Transmission):* For the enhanced Poisson $i\mathcal{S}$-graph $G_L$ with $L$ sectors, the out-degree $N_{\text{out}}$ of a typical node has the following negative binomial PMF

$$p_{N_{\text{out}}}(n) = \binom{L+n-1}{L-1} \left(\frac{\lambda_\ell}{\lambda_\ell + \lambda_e}\right)^n \left(\frac{\lambda_e}{\lambda_\ell + \lambda_e}\right)^L \quad (4)$$

for $n \geq 0$. In particular, the average node degrees are given by

$$\mathbb{E}\{N_{\text{out}}\} = \mathbb{E}\{N_{\text{in}}\} = L\frac{\lambda_\ell}{\lambda_e}. \quad (5)$$

*Proof:* We consider the process $\Pi_\ell \cup \{0\}$ obtained by adding a legitimate node to the origin of the coordinate system, and denote the out-degree of node 0 by $N_{\text{out}}$, as depicted in Fig. 2. Let $\mathcal{S}^{(l)}$ denote the $l$-th sector of node 0, where we omitted the subscript 0 for simplicity. Let $\{R_{\ell,i}^{(l)}\}_{i=1}^{\infty}$ be the distances—not necessarily ordered—between the origin and the legitimate nodes falling inside $\mathcal{S}^{(l)}$ (we similarly define $\{R_{e,i}^{(l)}\}_{i=1}^{\infty}$ for the eavesdroppers falling inside $\mathcal{S}^{(l)}$). Then, $N_{\text{out}} = \sum_{l=1}^{L} N_{\text{out}}^{(l)}$, where $N_{\text{out}}^{(l)} \triangleq \#\{R_{\ell,i}^{(l)} : R_{\ell,i}^{(l)} < \min_k R_{e,k}^{(l)}\}$ is the out-degree of node 0 associated with sector $l$. Furthermore, the random variables (RVs) $\{N_{\text{out}}^{(l)}\}$ are independent identically distributed (IID) for different $l$.[6] To determine the PMF of $N_{\text{out}}^{(l)}$, we use the fact that $\{(R_{\ell,i}^{(l)})^2\}_{i=1}^{\infty}$ and $\{(R_{e,i}^{(l)})^2\}_{i=1}^{\infty}$ are homogeneous Poisson processes with rates $\frac{\pi \lambda_\ell}{L}$ and $\frac{\pi \lambda_e}{L}$, respectively (by the mapping theorem [35, Sec. 2.3]). Following steps analogous to the proof of [30, Thm. 3.1], it is easy to show that $N_{\text{out}}^{(l)}$ has a geometric PMF given by $p_{N_{\text{out}}^{(l)}}(n) = p^n(1-p)$, $n \geq 0$, with $p = \frac{\lambda_\ell}{\lambda_\ell + \lambda_e}$. Now, since the RVs $\{N_{\text{out}}^{(l)}\}$ are IID in $l$, the total out-degree $N_{\text{out}}$ has a negative binomial PMF with $L$ degrees of freedom and the same parameter $p$, i.e., $p_{N_{\text{out}}}(n) = \binom{L+n-1}{L-1} p^n (1-p)^L$, $n \geq 0$, with $p = \frac{\lambda_\ell}{\lambda_\ell + \lambda_e}$. This is the result in (4), from which (5) follows trivially. □

We conclude that the average node degree increases *linearly* with the number of sectors $L$, and hence sectorized transmission is an effective technique for enhancing the secrecy of communications. Figure 2 provides an intuitive understanding of why sectorization works. Specifically, if there was no sectorization, node 0 would be out-isolated, due to the close proximity of the eavesdropper in sector $\mathcal{S}^{(4)}$. However, if we allow independent transmissions in four non-overlapping sectors, that same eavesdropper can only hear the transmissions inside sector $\mathcal{S}^{(4)}$. Thus, even though node 0 is out-isolated with respect to sector $\mathcal{S}^{(4)}$, it may still communicate securely with some legitimate nodes inside sectors $\mathcal{S}^{(1)}$, $\mathcal{S}^{(2)}$, and $\mathcal{S}^{(3)}$. Lastly, note that for $L = 1$ (i.e., no enhancement), Theorem 3.2 reduces to Theorem 3.1 as expected.

### C. Secure Communication With Eavesdropper Neutralization

In some scenarios, each legitimate node may be able to physically inspect its surroundings and deactivate the eavesdroppers falling inside some neutralization region. Our goal is to characterize the impact of such region on the local connectivity of the resulting $i\mathcal{S}$-graph.

With each node $x_i \in \Pi_\ell$, we associate a *neutralization region* $\Theta_i$ inside which all eavesdroppers have been deactivated. The *total neutralization region* $\Theta$ can then be seen as a Boolean model with points $\{x_i\}$ and associated sets $\{\Theta_i\}$, i.e., [37]

$$\Theta = \bigcup_{i=1}^{\infty} (x_i + \Theta_i).$$

Since the eavesdroppers inside $\Theta$ have been deactivated, the *effective eavesdropper process* after neutralization is $\Pi_e \cap \overline{\Theta}$, where $\overline{\Theta} \triangleq \mathbb{R}^2 \setminus \Theta$ denotes the complement of $\Theta$. The resulting $i\mathcal{S}$-graph with neutralization, $G_\Theta = \{\Pi_\ell, \mathcal{E}_\Theta\}$, has an edge set given by

$$\mathcal{E}_\Theta = \left\{\overrightarrow{x_i x_j} : |x_i - x_j| < |x_i - e^*|, \quad e^* = \operatorname*{argmin}_{e_k \in \Pi_e \cap \overline{\Theta}} |x_i - e_k|\right\}$$

i.e., the secure link $\overrightarrow{x_i x_j}$ exists if and only if $x_j$ is closer to $x_i$ than any other eavesdropper that has not been neutralized. In the following, we consider the case of a circular neutralization set, i.e., $\Theta_i = \mathcal{B}_0(\rho)$, where $\rho$ is a deterministic *neutralization radius*, and denote the corresponding graph by $G_\rho$.[7]

*Theorem 3.3 (Eavesdropper Neutralization):* For the enhanced Poisson $i\mathcal{S}$-graph $G_\rho$ with neutralization radius $\rho$,

---

[6]This is because the sectors $\{\mathcal{S}^{(l)}\}_{l=1}^{L}$ correspond to an equipartition of $\mathbb{R}^2$, and the processes $\Pi_\ell, \Pi_e$ are homogeneous Poisson.

[7]We use $\mathcal{B}_x(\rho) \triangleq \{y \in \mathbb{R}^2 : |y - x| \leq \rho\}$ to denote the closed two-dimensional ball centered at point $x$, with radius $\rho$.

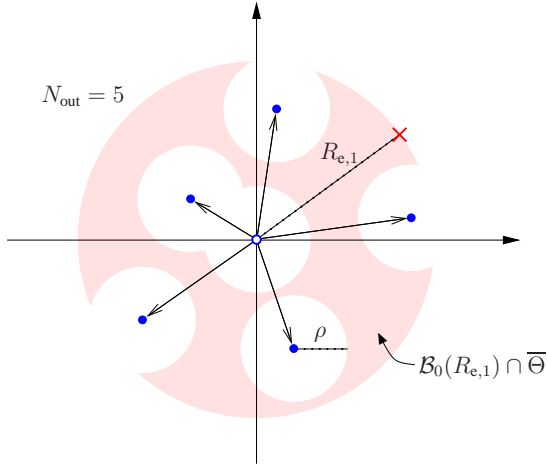

Figure 3. Secure communication with eavesdropper neutralization. In this example, the node at the origin can transmit messages with information-theoretic security to $N_{\text{out}} = 5$ nodes.

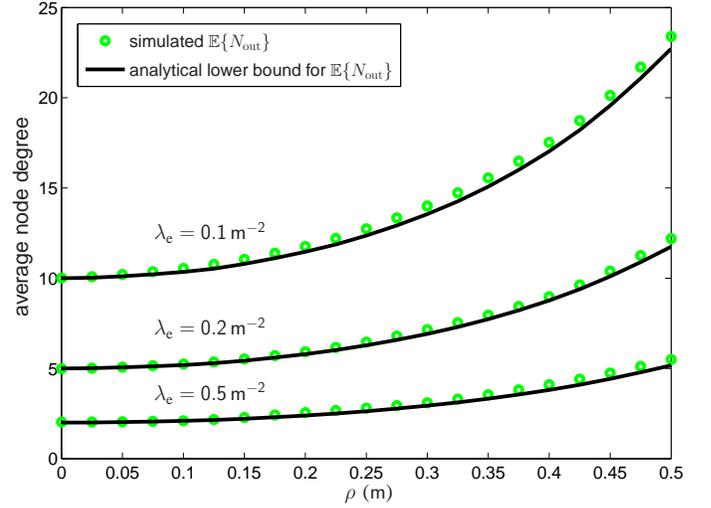

Figure 4. Average node degree versus the neutralization radius $\rho$, for various values of $\lambda_{\text{e}}$ ($\lambda_\ell = 1\,\text{m}^{-2}$).

the average node degrees of a typical node are lower-bounded by

$$\mathbb{E}\{N_{\text{out}}\} = \mathbb{E}\{N_{\text{in}}\} \geq \frac{\lambda_\ell}{\lambda_{\text{e}}}\left(\pi \lambda_{\text{e}} \rho^2 + e^{\pi \lambda_\ell \rho^2}\right). \quad (6)$$

*Proof:* We consider the process $\Pi_\ell \cup \{0\}$ obtained by adding a legitimate node to the origin of the coordinate system, and denote the out-degree of node 0 by $N_{\text{out}}$, as depicted in Fig. 3. Let $R_{\text{e},1} \triangleq \min_{e_k \in \Pi_{\text{e}} \cap \overline{\Theta}} |e_k|$ be the random distance between the first non-neutralized eavesdropper and the origin. Noting that

$$N_{\text{out}} = \sum_{x_i \in \Pi_\ell} \mathbb{1}\{|x_i| < R_{\text{e},1}\} = \iint_{\mathbb{R}^2} \mathbb{1}\{|x| < R_{\text{e},1}\} \Pi_\ell(dx),$$

we can use Fubini's theorem to write

$$\mathbb{E}\{N_{\text{out}}\} = \lambda_\ell \iint_{\mathbb{R}^2} \mathbb{P}_x\{|x| < R_{\text{e},1}\} dx$$

$$= \lambda_\ell \pi \rho^2 + \lambda_\ell \iint_{\mathcal{D}(\rho,\infty)} \mathbb{P}_x\{|x| < R_{\text{e},1}\} dx, \quad (7)$$

where $\mathcal{D}(a,b) \triangleq \{x \in \mathbb{R}^2 : a \leq |x| \leq b\}$ denotes the annulus centered at the origin, with inner radius $a$ and outer radius $b$; and $\mathbb{P}_x\{\cdot\}$ is the Palm probability associated with point $x$ of process $\Pi_\ell$.[8] Appendix A shows that the integrand above satisfies

$$\mathbb{P}_x\{|x| < R_{\text{e},1}\} \geq \exp\left(-\pi \lambda_{\text{e}} e^{-\lambda_\ell \pi \rho^2}(|x|^2 - \rho^2)\right). \quad (8)$$

Replacing (8) into (7), we have

$$\mathbb{E}\{N_{\text{out}}\} \geq \lambda_\ell \pi \rho^2 + \lambda_\ell \iint_{\mathcal{D}(\rho,\infty)} \exp\left(-\pi \lambda_{\text{e}} e^{-\lambda_\ell \pi \rho^2}(|x|^2 - \rho^2)\right) dx$$

$$= \lambda_\ell \pi \rho^2 + \frac{\lambda_\ell}{\lambda_{\text{e}}} e^{\lambda_\ell \pi \rho^2}.$$

Rearranging terms and noting that $\mathbb{E}\{N_{\text{out}}\} = \mathbb{E}\{N_{\text{in}}\}$, we obtain (6). □

[8] Informally, the Palm probability $\mathbb{P}_x\{\cdot\}$ can be interpreted as the conditional probability $\mathbb{P}\{\cdot | x \in \Pi_\ell\}$. Since the conditioning event has probability zero, such conditional probability is ambiguous without further explanation. Palm theory makes this notion mathematically precise (see [37, Sec. 4.4] for a detailed treatment).

We conclude that the average node degree increases at a rate that is at least *exponential* with the neutralization radius $\rho$, making eavesdropper neutralization an effective technique for enhancing the secrecy of communications.

## IV. NUMERICAL RESULTS

Figure 4 illustrates the effectiveness of eavesdropper neutralization in enhancing secure connectivity. In particular, it plots the average node degree versus the neutralization radius $\rho$, for various values of $\lambda_{\text{e}}$. We observe that the analytical lower-bound for $\mathbb{E}\{N_{\text{out}}\}$ given in (6) is very close to the actual value of $\mathbb{E}\{N_{\text{out}}\}$ obtained through Monte Carlo simulation. The lower-bound becomes tight in the following two extreme cases:

1) $\rho = 0$: This corresponds to the case of no enhancement, so from (3) we have $\mathbb{E}\{N_{\text{out}}\} = \frac{\lambda_\ell}{\lambda_{\text{e}}}$. Since the bound in (6) also equals $\frac{\lambda_\ell}{\lambda_{\text{e}}}$ for $\rho = 0$, it is tight.
2) $\lambda_{\text{e}} \to \infty$: In the limit, at least one eavesdropper will fall almost surely inside the annulus $\mathcal{D}(\rho, \rho + \epsilon)$, for any $\epsilon > 0$. As a result, $\mathbb{E}\{N_{\text{out}}\}$ approaches the average number of legitimate nodes inside the ball $\mathcal{B}_0(\rho)$, i.e., $\lambda_\ell \pi \rho^2$. Since the bound in (6) also approaches $\lambda_\ell \pi \rho^2$ as $\lambda_{\text{e}} \to \infty$, it is asymptotically tight.

## V. CONCLUSIONS

In this paper, we proposed two techniques shown to dramatically enhance the connectivity of the $i\mathcal{S}$-graph: *sectorized transmission* and *eavesdropper neutralization*. We proved that if each legitimate node is able to transmit independently in $L$ sectors of the plane, the average node degree increases *linearly* with $L$. On the other hand, if each legitimate node is able to neutralize all eavesdroppers within a radius $\rho$, the average degree increases *at least exponentially* with $\rho$. We are hopeful that further efforts will provide an understanding of how *long-range* secure communication is improved by the two proposed strategies. For example, the effect of $L$ and $\rho$ on the critical densities for continuum percolation is still unknown, as is their effect on the secrecy capacity scaling laws.

## APPENDIX A
## DERIVATION OF (8)

Because $\Pi_\ell$ is a Poisson process, the Palm probability $\mathbb{P}_x\{|x| < R_{e,1}\}$ in (7) can be computed using Slivnyak's theorem by adding a legitimate node at location $x$ to $\Pi_\ell$. For a fixed $x \in \mathcal{D}(\rho, \infty)$, we can thus write

$$\mathbb{P}_x\{|x| < R_{e,1}\} = \mathbb{P}_{\Theta,\Pi_e}\{\Pi_e\{\overline{\Theta} \cap \mathcal{D}(\rho, |x|) \setminus \mathcal{B}_x(\rho)\} = 0\}$$
$$\geq \mathbb{P}_{\Theta,\Pi_e}\{\Pi_e\{\overline{\Theta} \cap \mathcal{D}(\rho, |x|)\} = 0\}$$
$$= \mathbb{E}_\Theta\{\exp(-\lambda_e \mathbb{A}\{\overline{\Theta} \cap \mathcal{D}(\rho, |x|)\})\} \quad (9)$$
$$\geq \exp(-\lambda_e \mathbb{E}_\Theta\{\mathbb{A}\{\overline{\Theta} \cap \mathcal{D}(\rho, |x|)\}\}), \quad (10)$$

where $\mathbb{A}\{\mathcal{R}\}$ to denotes the area of the arbitrary region $\mathcal{R}$. Equation (9) follows from conditioning on $\Theta$, and using the fact that $\Pi_e$ and $\Theta$ are independent. Equation (10) follows from Jensen's inequality. The term inside the exponential in (10) corresponds to the average area of a random shape, and can be computed using Fubini's theorem as

$$\mathbb{E}_\Theta\{\mathbb{A}\{\overline{\Theta} \cap \mathcal{D}(\rho, |x|)\}\}$$
$$= \mathbb{E}_\Theta\left\{\iint_{\mathbb{R}^2} \mathbb{1}\{y \in \overline{\Theta} \cap \mathcal{D}(\rho, |x|)\} dy\right\}$$
$$= \iint_{\mathcal{D}(\rho, |x|)} \mathbb{P}\{y \in \overline{\Theta}\} dy$$
$$= \iint_{\mathcal{D}(\rho, |x|)} \mathbb{P}\{\Pi_\ell\{\mathcal{B}_y(\rho)\} = 0\} dy$$
$$= \iint_{\mathcal{D}(\rho, |x|)} e^{-\lambda_\ell \pi \rho^2} dy$$
$$= \pi(|x|^2 - \rho^2) e^{-\lambda_\ell \pi \rho^2}. \quad (11)$$

Replacing (11) into (10), we obtain the desired inequality in (8).